\documentclass[aps,11pt,final,notitlepage,oneside,onecolumn,nobibnotes,nofootinbib,preprintnumbers,superscriptaddress,showpacs,centertags,showkeys,amsmath,amssymb]{revtex4}

\usepackage{graphicx} 
\usepackage{dcolumn} 

\def\figurename{Fig.}

\newcommand{\RR}{\mathrm {Re}}
\newcommand{\II}{\mathrm {Im}}

\def\i{\mathrm i}
\def\phi{\varphi}

\begin{document}
\begin{flushright}
{LNF - 08 / 17(P)} \\
{July 14, 2008}\\
\end{flushright}

\title{QCD Contributions to the Froissart bound for the total cross-section}

\author{A. Achilli}
\email{achilli@fisica.unipg.it}
\affiliation{INFN and Physics Department University of Perugia, I-06123 Perugia, Italy}

\author{A. Grau}
\email{igrau@ugr.es}
\affiliation{Departamento de Fisica Te\'orica y del Cosmos
Universidad de Granada, 18071 Granada, Spain}

\author{G. Pancheri}
\email{pancheri@lnf.infn.it}
\affiliation{INFN Frascati National Laboratories, I-00044 Frascati, Italy}

\author{Y. N. Srivastava}
\email{Yogendra.Srivastava@pg.infn.it}
\affiliation{INFN and Physics Department University of Perugia, I-06123 Perugia, Italy}

\date{\today}

\begin{abstract}
We discuss the effect of infrared soft gluons on the asymptotic behaviour of the total cross-section. We use a singular but integrable expression for the strong coupling constant in the infrared limit and relate its
behaviour to the satisfaction of the Froissart bound, giving a specific phenomenological example.

\end{abstract}


\maketitle

\begin{center}
{\it Presented at Hadron Structure '07, HS07, Modra-Harm\'onia, September 3-7, 2007}
\end{center}

\section{Introduction}
The understanding of the energy behaviour of total cross-sections in hadron and photon scattering is still  an unresolved challenge. In our opinion, the major problem can be summarized as follows. The simplest and most successful parametrization of the total cross-section for proton-proton and proton-antiproton scattering  \cite{dl}
follows from the optical theorem and describes the initial fall through the exchange of  Regge trajectories, i.e. $s^{\alpha_R(0)-1}$  with $\alpha_R(0)<1$ and the rise as due to  Pomeron
exchange, i.e. $ s^{\alpha_P(0)-1}$ with $\alpha_P(0)>1$. Although successful, this parametrization violates the Froissart bound on total cross-sections, established long time ago \cite{froissart}, namely that $\sigma^{tot}\le \log^2{s}$ as $s \rightarrow \infty$. Other parametrizations have then been proposed, which are based on purely phenomenological grounds, but as such  do not really give an insight into the mechanisms of the rise of total cross-sections. In another vein, there are the eikonal models which use the impact parameter distribution of protons, often derived from the  Fourier trasform of the proton form factor, and a basic scattering  cross-sections \cite{durand}. Among them is the eikonal mini-jet model in which the term which drives the rise is a perturbatively calculated QCD jet cross-section . Our proposal is to use this eikonal mini-jet model with an impact parameter factor derived from the Fourier transform of the initial  transverse momentum distribution of valence quarks in the proton, calculated through soft gluon resummation. This distribution is energy dependent and produces a saturation mechanism which quenches the mini-jet rise in such a way that one restores the Froissart bound. In the following we shall first describe the soft gluon transverse momentum distribution, then the eikonal mini-jet model and its application to the proton cross-sections. We shall then show how this model for total cross-sections, in the very large asymptotic regime, and with a particular ans\"atz for the zero momentum modes of the emitted soft gluons,  satisfies the Froissart bound.

\section{The soft gluon transverse momentum distribution}
Soft gluon emission from the initial quarks and gluons introduces
an acollinearity between the initial partons. Such acollinearity
is energy dependent, and can be responsible for reducing the fast
rise of the mini-jet cross-section \cite{corsetti}. We have
described this effect in a number of papers, here we first discuss
in detail the main characteristics of the soft gluon resummed
transverse momentum distribution.

The emission of an infinite number of low momentum massless
quanta  results in a four momentum distribution
\begin{equation}
d^4P(K)={{d^4K}\over{(2\pi)^4}} \int d^4 x e^{iK\cdot x} e^{-h(x)}
\label{d4p}
\end{equation}
with
\begin{equation}
h(x)=\int d^3{\bar n}(k) [1-e^{-ik\cdot x}]
\end{equation}
where $d^3{\bar n}(k)$ is the average number of soft quanta,
gluons or photons, emitted by a color or electrically charged
source. Upon integrating on the energy and longitudinal momentum
variable, one obtains the well known transverse momentum
distribution
  \begin{equation}
d^2P(K_{\perp})={{d^2{\bf K}_{\perp}}\over{(2\pi)^2}} \int d^2 {\bf b} e^{i{\bf K_{\perp}\cdot b}} e^{-h(b)}
\label{d2p}
\end{equation}
with
\begin{equation}
h(b)=\int d^3{\bar n}(k) [1-e^{-i{\bf k_t\cdot b}}]
\label{softterm}
\end{equation}
The above transverse momentum distribution does not admit a closed
form solution, unlike    the energy distribution in QED, which easily
integrates to a power law function. This has led to various
strategies to evaluate the integral in Eq. (\ref{d2p}), none of
which is however fully satisfying, for reasons which we shall try
to illustrate. Indeed, while the importance of this function has
been known for a long time, in our opinion such importance has not
been fully exploited, and in  this note we shall discuss this point and present our approach to the problem.

The simplest case to examine is that of  the low momentum  soft
quanta  in an Abelian gauge theory with a constant but not small
coupling constant. It is then possible to make  some simple
approximations, as in \cite{oldkt}, and
obtain a closed form for the transverse momentum distribution, namely
\begin{equation}
d^2 P(K_ \bot  ) = \frac{{\beta (2\pi )^{ - 1} }} {{\Gamma (1 +
\beta /2)}}\frac{{d^2 K_ \bot  }} {{2E^2 A}}\left( {\frac{{K_ \bot
}} {{2E\sqrt A }}} \right)^{\frac{\beta } {2} - 1}  \times
\mathcal{K} _{1 - \beta /2} \left( {\frac{{K_ \bot  }} {{E\sqrt A
}}} \right)
\end{equation}
where ${\cal K}_{1-\beta/2}$ is the modified Bessel function of
third kind, $\beta$ is obtained by performing the angular integration over $d^3{\bar n}(k)$, E is the maximum energy allowed to the single soft
quantum emitted and A is related to $<K_ \bot>$ and defined so that the
distribution is normalized to 1.

The above procedure fails in the non Abelian case, where
the strong coupling constant is large but  not  constant.
This was  included \cite{ddt,pp,collins} in the
QCD versions of the transverse momentum distribution, still
presently used in many phenomenological applications. Using the
asymptotic freedom expression for $\alpha_s$,  the relative  transverse momentum
distribution induced by soft gluon emission from  a pair of,
initially collinear, colliding partons  is  obtained, at LO, as
\begin{equation}
\begin{centering}
h(b,E) = \frac{16}{3} \int_\mu^E {\alpha_s(k_t)\over{\pi}}
 {{dk_t}\over{k_t}}
  \ln {
  {{2E}\over{k_t}}}
  \approx \frac{32}{33-2N_f}\left\{ {\ln ( {\frac{{2E}}
{\Lambda }} )\left[ {\ln ({\ln ( {\frac{E} {\Lambda }})}) - \ln
({\ln ( {\frac{\mu } {\Lambda }} )})} \right] - \ln ( {\frac{E}
{\mu }} )} \right\}
\end{centering}
\end{equation} where the
integration only extends down to a scale $\mu$. Because of the cut-off $\mu$, this expression
fails to reproduce the entire range of low energy transverse
momentum effects and an {\it ad hoc} constant,  i.e. the intrinsic transverse momentum, is
introduced for phenomenological applications.

The problem we observe in the above use of this distribution is
its limited applicability to minimum bias physics and its lack of
connection to the question of confinement.  Our  approach instead is to extend the integral over the soft gluon momentum down to zero modes; propose an ans\"atz for the behaviour of $\alpha_s$ in the infrared region, and compare its consequences  on measurable quantities such as the total cross-section.
In the next section, we shall present one such  phenomenological application
through the calculation of total cross-sections  and in the last section
 discuss the consequences of our ans\"atz for the infrared soft gluon limit to the
limitations imposed by the Froissart bound on the total
cross-section.

\section{A model for Total Cross-section}
Recently we have obtained predictions for the total cross-section
\cite{ourlast,lastPLB,PRD60} at LHC energies using a model based on hard
component of scattering responsible for the rise of the total
cross-section and soft gluon emission from scattering particles
which softens the rise and gives the impact parameter
distribution. In this model mini-jet events coming from parton-parton high energy collisions
are responsible for the rise of the total
cross-section. From perturbative QCD, we have the following
expression for parton-parton cross section at high energy
\begin{equation}
 \sigma^{AB}_{\rm jet} (s,p_{tmin})=
\int_{p_{tmin}}^{\sqrt{s/2}} d p_t \int_{4
p_t^2/s}^1 d x_1  \int_{4 p_t^2/(x_1 s)}^1 d x_2 \times
\sum_{i,j,k,l}
f_{i|A}(x_1,p_t^2) f_{j|B}(x_2,p_t^2)
  \frac { d \hat{\sigma}_{ij}^{ kl}(\hat{s})} {d p_t}
\label{sjet}
\end{equation}
which depends on the minimum transverse momentum allowed to the
scattered partons in the final state,  $p_{tmin}$,  $\approx 1-2$ GeV. This parameter represents the energy scale
which distinguishes hard processes from the soft ones.
Evaluating the mini-jet cross-sections using different LO
parametrizations available for the Partonic Density Functions
$f_{i|A}$ \cite{densities},  we obtain a growth of $\sigma_{jet}$
with energy as a (small) power of s. Soft gluon emission changes the parton
collinearity and contributes in softening the rise of the total cross
section. The average number of soft gluon emissions increases with
energy, with  more emission leading to more acollinearity between colliding
partons and less energy available for mini-jet production. These
processes are best studied through an impact parameter (b-)distribution. The
overlap function which describes the b-distribution of the incident
partons in our model is
\begin{equation}
A_{BN}(b,s)=N \int d^2{\bf  K_{\perp}}\  e^{-i{\bf K_\perp\cdot b}}
 {{d^2P({\bf K_\perp})}\over{d^2 {\bf K_\perp}}}={{e^{-h( b,q_{max})}}\over
 {\int d^2{\bf b} \ e^{-h( b,q_{max})}
 }}
\label{overlap}
\end{equation}
with
\begin{equation}
h( b,q_{max}) =  \frac{16}{3}\int_0^{q_{max} }
 {{ \alpha_s(k_t^2) }\over{\pi}}{{d
 k_t}\over{k_t}}\log{{2q_{\max}}\over{k_t}}[1-J_0(k_tb)]
\end{equation}
similar to the one obtained in the first section, but for the fact that the integral extends down to zero momentum and $J_0(k_tb)$ needs to be retained for infrared finiteness.
For $\alpha_s$ we use
a phenomenological expression, which coincides with the usual QCD limit for large $k_t$, but is  singular
for $k_t \to 0$, namely
\begin{equation}
\alpha_s(k_t^2)={{12 \pi}\over{33-2N_f}} {{p}
\over{\ln[1+p({{k_t}\over{\Lambda}})^{2p}]}}
\end{equation}
with $p<1$ for the integral over $k_t$ to exist and $p \geq 1/2$ for the
correct analyticity in the momentum transfer variable.

The parameter $q_{max}$ is linked by kinematics to the maximum transverse
momentum allowed to single soft gluon emission in a
given hard collision, and for this purpose, we use an expression averaged over
all the parton densities\cite{PRD60}
\begin{equation}
q_{\max } (s)   = \sqrt {\frac{s} {2}}\,
\frac{{\sum\limits_{i,j} {\int {\frac{{dx_1 }} {{x_1 }}\int
{\frac{{dx_2 }} {{x_2 }} \int_{z_{min}}^1 {dz f_i (x_1) f_j (x_2)
\sqrt {x_1 x_2 } (1 - z)} } } } }} {{\sum\limits_{i,j} {\int
{\frac{{dx_1 }} {{x_1 }}\int {\frac{{dx_2}} {{x_2}}
\int_{z_{min}}^1 {dz} f_i (x_1)f_j (x_2) } } } }}
\end{equation}
with $z_{min}=4p_{tmin}^2/(s x_1 x_2)$.

The use of eikonal representation allows us to consider
multiple scatterings. Neglecting the real part of the eikonal
function, the expression for the total cross section is
given by
\begin{equation}
\sigma_{tot}=2\int d^2{\bf b}[1-e^{-n(b,s)/2}]
\end{equation}
with $n(b,s)=n_{soft}(b,s) +
n_{hard}(b,s)=n_{soft}(b,s)+A_{BN}(b,s)\sigma_{jet}(s,p_{tmin})$
denoting the average number of partonic collisions during the scattering.
The hard term is evaluated using the expressions (\ref{sjet}) and
(\ref{overlap}) for $\sigma_{jet}$ and $A_{BN}$, while the soft
one has a phenomenological parametrization
\begin{equation}
n_{soft}(b,s) = A _{BN} ^{soft} (b,s) \sigma_0 ( 1 + \epsilon \frac{2}{\sqrt{s}})
\end{equation}
where $\sigma_0=constant$ is a parameter necessary to reproduce the
right normalization of the total cross-section, while
$\epsilon=0,1$ respectively for $pp$ and $p\bar{p}$ scattering, $A _{BN} ^{soft}$ parametrized as
in  \cite{ourlast}.

\begin{figure} [hbt!]
\includegraphics[height=3.5in]{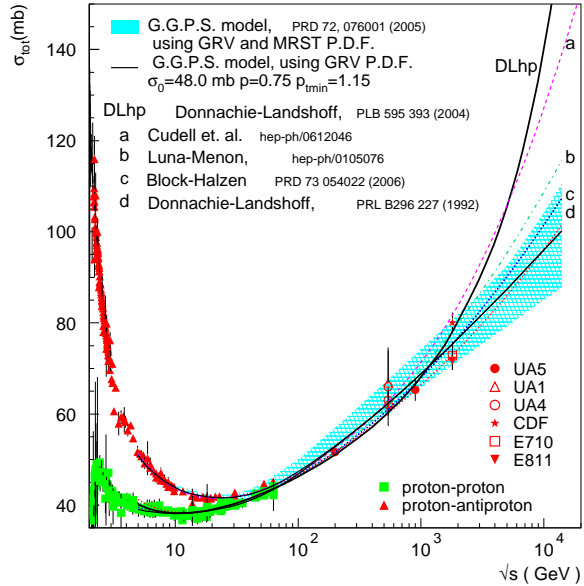}
\caption{\label{fig:1} Total cross-section obtained from our model \cite{PRD60}
using different PDF's \cite{densities}, compared with data
\cite{data} and with other models \cite{bibicross}}
\end{figure}

\figurename{\ref{fig:1}} from \cite{lastPLB} shows the range of values of the total
cross-section produced with this model using a set of
phenomenological values for the parameters $p_{tmin}$, $\sigma_0$
and $p$, and varying the parton densities.

\section{Soft gluons in the infrared limit and total cross-section}
At very large energies, using the eikonal representation,  the total cross-section in our model \cite{ourlast} reads
\begin{equation}
  \sigma _T (s) \approx 2\pi \int_0^\infty  {db^2 } [1 - e^{ - n_{hard} (b,s)/2} ]
\end{equation}
where $n_{hard} (b,s) = \sigma _{jet} (s)A_{hard} (b,s)$. We
consider the asymptotic expression for $\sigma_{jet}$ at high
energies, which grows as a power of $s$: $\sigma _{jet} (s)
\approx \sigma _1 (s/s_0 )^\varepsilon $. $A_{hard}(b,s)$ is obtained
through soft gluon resummation as
\begin{equation}
  A_{hard} (b,s) \propto e^{ - h(b,s)}
\end{equation}
where $h(b,s)$\ is given by Eq. (\ref{softterm}). If we take now
the limit $k_t \to 0$ where $d^3 n (k) \propto \alpha _s (k_t ^2
)$ and $\alpha _s (k_t ) \approx (\frac{\Lambda}{k_t})^{2p}$, with
$p<1$, the integral (\ref{softterm}) gives $h(b,s) \propto (b
{\bar \Lambda})^{2p}$. This  implies  $A_{hard} (b) \propto e^{ -
(b{\bar \Lambda})^{2p} }$ and
\begin{equation}
  n_{hard}  = 2C(s)e^{ - (b{\bar \Lambda})^{2p} }
\end{equation}
where  $C(s) = A_0 (s/s_0 )^\varepsilon  \sigma _1$. The resulting
expression for $\sigma_T$ is
\begin{equation}
  \sigma _T (s) \approx 2\pi \int_0^\infty  {db^2 } [1 - e^{ -
C(s)e^{ - (b{\bar \Lambda})^{2p} } } ]
\end{equation}
from which we derive
\begin{equation}
  \sigma _T  \to [\varepsilon \ln (s)]^{(1/p)}
\end{equation}
The result shows explicitly that soft gluon resummation can restore the asymptotic
growth of $\sigma_T$ to lie within the Froissart bound. (Recall that $1/2\ \leq p < 1$).
\section{Conclusions}
We have shown how soft gluon resummation acts as a saturation mechanism in the rise of the total cross-section. This effect, embedded into the eikonal formalism with QCD mini-jets to drive the rise, leads to asymptotic cross-sections which satisfy the Froissart bound.

\end{document}